\documentclass[english]{article}
\usepackage[T1]{fontenc}
\usepackage[latin1]{inputenc}
\usepackage{babel}

\makeatletter

\providecommand{\LyX}{L\kern-.1667em\lower.25em\hbox{Y}\kern-.125emX\@}

 \newcommand{\lyxaddress}[1]{
   \par {\raggedright #1 
   \vspace{1.4em}
   \noindent\par}
 }

\usepackage[T1]{fontenc}
\usepackage{graphics}


\begin{document}

\title{Magnetization of coupled spin clusters in Ladder Geometry}

\author{Emily Chattopadhyay and Indrani Bose}

\maketitle

\lyxaddress{\centering \textbf{Physics Department, Bose Institute, 93/1, A.P.C.
Road, Calcutta-700009, India}}

\begin{abstract}
In this paper, we construct a class of spin-1/2 antiferromagnetic
(AFM) two-chain ladder models consisting of blocks of n-spin tetrahedral
clusters alternating with two-spin rungs. For n=4 and 6 and in extended
parameter regimes, the exact ground state of the ladder is shown to
be a product of the ground states of the rungs and the n-spin blocks,
in both zero and finite magnetic fields. In the latter case, magnetization/site
\( m \) versus magenetic field \( h \) plot exhibits well-defined
magnetization plateaus.\\
 PACS numbers: 75.10 Jm, 75.40 Mg, 75.50Ee 
\end{abstract}
Spin ladders have been widely studied in recent times as these systems
exhibit a variety of novel phenomena in the undoped as well as the
doped states\( ^{1,2} \). There are also several magnetic compounds
which can be considered as coupled spin cluster systems in which the
dominant exchange interactions occur within clusters of spins. The
clusters are coupled through weaker exchange interactions. A prominent
example of such systems is that of molecular magnets\( ^{3} \). Examples
of spin clusters are dimers and four-spin plaquettes. The magnetic
properties of spin clusters can be determined exactly if the size
of the cluster is small. It is of significant interest to find out
how the cluster properties are modified in the bulk. In this paper,
we propose a class of two-chain spin ladder models which can be defined
in terms of n-spin clusters and dimers. The specific values of n considered
are n = 4 and 6 though a generalisation to higher n values is also
possible. The models describe spins of magnitude 1/2 interacting via
modulated antiferromagnetic (AFM) exchange interactions. We show that
in a wide parameter regime, the exact ground and low-lying excited
states of the full ladder model can be described in terms of the eigenstates
of the dimers and n-spin clusters, i.e., the clusters act as decoupled
entities even in the bulk. This is also true when the exchange interactions
coupling the clusters are of considerable strength. In the same parameter
region, the magnetization/site of the ladder in the presence of an
external magnetic field exhibits the phenomenon of magnetization plateaus.
The condition for the appearance of a plateau is given by\( ^{4} \)

\begin{equation}
\label{1}
S_{u}-m_{u}=integer
\end{equation}
 where \( S_{u} \) and \( m_{u} \) are the total spin and magnetization
in unit period of the ground state.

Spin-1/2 ladder models of various types have been extensively studied
both in zero and finite magnetic fields\( ^{5-16} \). In frustrated
spin ladder models, extra diagonal exchange couplings (one or two)
are present. Hakobyan et al.\( ^{10} \) have given an overview on
the phase diagram of the general frustrated two-chain ladder model.
The magnetization process of the general model is, however, yet to
be undertaken. Brenig et al.\( ^{11} \)have introduced a dimerized
and completely frustrated two-chain ladder model. The model, in which
diagonal exchange interactions are present in every plaquette, is
equivalent to a chain of edge sharing tetrahedra. The rung exchange
interactions in the ladder model are of strength \( J_{2} \). The
nearest -neighbour intra-chain and diagonal exchange interactions
are of equal strengths and in alternate plaquettes the strengths are
\( J_{1} \) and \( J_{3} \) respectively. Examples of tetrahedral
cluster compounds are tellurates of the type \( Ca_{2}Te_{2}O_{5}X_{2} \)
with \( X=Cl \), \( Br \)\( ^{12} \). The class of models we propose
describes two-chain spin-ladders with modulated exchange interactions.
The spin ladders consist of tetrahedral spin clusters containing n
spins separated by two-spin clusters, i.e., dimers (rungs). In section
II, we introduce the models and discuss the magnetization properties
in the presence of an external magnetic field. Section III contains
a summary and discussion of the major results obtained.

The spin ladder model consists of blocks of tetrahedral clusters containing
n spins separated by rungs of two spins. Fig. 1. shows the simplest
such ladder model with n = 4. The tetrahedral clusters are represented
by solid lines and the two-spin rungs (dimers) by dashed lines. Within
a tetrahedral cluster, the diagonal exchange interactions are of strength
\( J_{3} \) and the other exchange interactions are of strength \( J_{1} \).
The rung exchange interaction strengths are \( J^{\prime } \) and
a rung is coupled to a neighbouring tetrahedral cluster through exchange
interactions (dashed lines) of strength \( J_{2} \). Periodic boundary
condition is assumed to hold true. The spin Hamiltonian describing
the ladder model is given by

\[
H=\sum _{i=3j+1,j=0,1,\cdots }[J_{1}(\overrightarrow{S}_{1i}.\overrightarrow{S}_{1i+1}+\overrightarrow{S}_{2i}.\overrightarrow{S}_{2i+1}+\overrightarrow{S}_{1i}.\overrightarrow{S}_{2i}+\overrightarrow{S}_{1i+1}.\overrightarrow{S}_{2i+1})\]

\[
+J_{3}(\overrightarrow{S}_{1i}.\overrightarrow{S}_{2i+1}+\overrightarrow{S}_{2i}.\overrightarrow{S}_{1i+1})]+J^{\prime }\sum _{i=3j,j=0,1,\cdots }\overrightarrow{S}_{1i}.\overrightarrow{S}_{2i}\]

\begin{equation}
\label{2}
+J_{2}\sum _{i=3j+2,j=0,1,\cdots }(\overrightarrow{S}_{1i}+\overrightarrow{S}_{2i}+\overrightarrow{S}_{1i+2}+\overrightarrow{S}_{2i+2}).(\overrightarrow{S}_{1i+1}+\overrightarrow{S}_{2i+1})
\end{equation}

\[
=H_{T}+H_{R}+H_{TR}\]

The spin operator \( \overrightarrow{S}_{1i} \) (\( \overrightarrow{S}_{2i} \)
) is associated with the \( i \)-th site of the lower (upper) chain
of the ladder, the site indices are sequential as shown in Fig. 1.
The Hamiltonians \( H_{T} \) and \( H_{R} \) describe the tetrahedral
clusters and the rungs respectively whereas \( H_{TR} \) contains
the exchange couplings between the tetrahedral clusters and the rungs.
The total spin of each rung is a conserved quantity due to the special
structure of the Hamiltonian.

We now determine the ground state of the ladder model. Using the method
of `divide and conquer'\( ^{17} \), it is easy to show that for \( J_{2}\leq \frac{J^{\prime }}{4} \),
the exact ground state has all the rungs and the tetrahedral clusters
in their ground state spin configurations. A brief sketch of the proof
is given in the following. The ground state of a rung is a singlet.
The ground state of a tetrahedral spin cluster is a resonating valence
bond (RVB) state and has total spin S=0. The ground state is \( \left| \psi _{RVB1}\right\rangle  \)
for \( J_{3}<J_{1} \) and \( \left| \psi _{RVB2}\right\rangle  \)
for\( J_{3}>J_{1} \). The states \( \left| \psi _{RVB1}\right\rangle  \)
and \( \left| \psi _{RVB2}\right\rangle  \) are linear combinations
(plus and minus) of two valence bond (VB) states. In one VB state,
the two VBs (singlets) are horizontal and in the other vertical.

{\centering \textbf{TABLE I}\par}

{\centering \begin{tabular}{|c|c|c|c|}
\hline 
S&
 Eigenvalues&
 S\( ^{z} \)&
 Eigenstates\\
\hline
\cline{4-4} 
0&
 \( -2J_{1}+\frac{J_{3}}{2} \)&
 0&
\multicolumn{1}{c|}{\( \left| \psi _{RVB1}\right\rangle  \)}\\
\hline
0&
 \( -\frac{3J_{3}}{2} \)&
 0&
\( \left| \psi _{RVB2}\right\rangle  \)\\
\hline
1&
 \( -J_{1}+ \)\( \frac{J_{3}}{2} \)&
 0&
 \( \uparrow \downarrow \downarrow \uparrow -\downarrow \uparrow \uparrow \downarrow  \)\\
\hline
1&
 \( -\frac{J_{3}}{2} \)&
 0&
 \( \uparrow \downarrow \uparrow \downarrow -\downarrow \uparrow \downarrow \uparrow  \)\\
\hline
1&
 \( -\frac{J_{3}}{2} \)&
 0&
 \( \uparrow \uparrow \downarrow \downarrow -\downarrow \downarrow \uparrow \uparrow  \)\\
\hline
2&
 \( J_{1}+\frac{J_{3}}{2} \)&
 0&
 \( \uparrow \uparrow \downarrow \downarrow +\downarrow \downarrow \uparrow \uparrow +\uparrow \downarrow \uparrow \downarrow +\downarrow \uparrow \downarrow \uparrow  \)\\
&
&
&
\( +\uparrow \downarrow \downarrow \uparrow +\downarrow \uparrow \uparrow \downarrow  \)\\
\hline
1&
 \( -\frac{J_{3}}{2} \)&
 1&
 \( \uparrow \uparrow \downarrow \uparrow -\uparrow \downarrow \uparrow \uparrow  \)\\
\hline
1&
 \( -J_{1}+ \)\( \frac{J_{3}}{2} \)&
 1&
 \( \uparrow \uparrow \uparrow \downarrow -\uparrow \uparrow \downarrow \uparrow -\uparrow \downarrow \uparrow \uparrow +\downarrow \uparrow \uparrow \uparrow  \)\\
\hline
1&
 \( J_{1}+\frac{J_{3}}{2} \)&
 1&
 \( \uparrow \uparrow \uparrow \downarrow +\downarrow \uparrow \uparrow \uparrow +\uparrow \uparrow \downarrow \uparrow +\uparrow \downarrow \uparrow \uparrow  \)\\
\hline
2&
 \( J_{1}+\frac{J_{3}}{2} \)&
 2&
 \( \uparrow \uparrow \uparrow \uparrow  \) \\
\hline
\end{tabular}\par}

Table I: The energy eigenvalues and eigenvectors of a tetrahedral
cluster with exchange interactions of strength \( J_{1} \) (horizontal
and vertical) and \( J_{3} \). The eigenstates \( \left| \psi _{RVB1}\right\rangle  \)
and \( \left| \psi _{RVB2}\right\rangle  \) are the resonating valence
bond states. \\
\vspace{.05cm}\\
At \( J_{3}=J_{1} \), the ground state of a tetrahedral cluster is
doubly degenerate. The two states have a pair of singlets (valence
bonds) along either the horizontal or the vertical bonds. The Hamiltonian
\( H_{TR} \) (Eq. (2)), containing the exchange couplings between
the rungs and the tetrahedral clusters, act on the rung singlets to
give zero and thus has no contributions to the energy \( E_{1} \)
of the eigenstate. \( E_{1} \) is the sum of the ground state energies
of the tetrahedral clusters and the rungs. Let \( E_{g} \) be the
exact ground state energy of the total Hamiltonian \( H \). Then
\( E_{g} \) is \( \leq E_{1} \). Let \( \left| \psi _{g}\right\rangle  \)
be the exact ground state wave function. Then, from variational theory,

\begin{equation}
\label{3}
E_{g}=\sum _{i}\left\langle \psi _{g}\right| H_{i}\left| \psi _{g}\right\rangle +\sum _{i}\left\langle \psi _{g}\right| H^{\prime }_{i}\left| \psi _{g}\right\rangle \geq \sum _{i}(E_{io}+E^{\prime }_{io})
\end{equation}

\[
H=\sum _{i}(H_{i}+H^{\prime }_{i})\]
 where \( H_{i} \) 's are the tetrahedral cluster Hamiltonians with
the ground state energy \( E_{io} \) (Table I) and \( H^{\prime }_{i} \)'s
are the six spin cluster Hamiltonians, each of which contains the
rung exchange interaction Hamiltonian and the eight exchange couplings
(four horizontal and four diagonal) which connect the rung to nearest-neighbour
tetrahedral clusters. The ground state energy of \( H^{\prime }_{i} \)
is \( E^{\prime }_{io} \). For \( J_{2}\leq \frac{J^{\prime }}{4} \),
\( E^{\prime }_{io} \) is the energy of a singlet across the rung.
We can now write down the inequality,

\begin{equation}
\label{4}
\sum _{i}(E_{io}+E^{\prime }_{io})\leq E_{g}\leq E_{1}
\end{equation}
 \( E_{1} \) is, however, exactly equal to \( \sum _{i}(E_{io}+E^{\prime }_{io}) \)
since it is the sum over the ground state energies of all the rungs
and the tetrahedral clusters. Thus, \( E_{g}=E_{1} \), i.e., the
exact eigenstate is also the exact ground state of the full ladder
model. The ground state has the novel structure of islands of four-spin
RVB configurations in the tetrahedral clusters separated by singlet
spin configurations along the rungs. The exact ground state energy
is \( E_{g}=N(E_{i0}-3\frac{J^{\prime }}{4}) \) where N is the total
number of tetrahedral clusters as well as rungs in the ladder. \( E_{i0}=-2J_{1}+\frac{J_{3}}{2} \)
for \( J_{3}<J_{1} \) and \( E_{i0}=-\frac{3J_{3}}{2} \) for \( J_{3}>J_{1} \)
(Table I). When \( J_{3}=J_{1} \), the exact ground state is highly
degenerate. The number of such states is \( 2^{N} \).

We now want to check whether the exact ground state is still a product
of the ground states of the rungs and the tetrahedral clusters when
\( J_{2} \) is made larger than \( \frac{J^{\prime }}{4} \). For
this, the total Hamiltonian \( H \) (Eq. (2)) is written as a sum
over six-spin sub-Hamiltonians, \( h_{i} \)'s, i.e., \( H=\sum _{i}h_{i} \).
Each sub-hamiltonian describes a tetrahedral cluster coupled to a
rung. The six-spin sub-Hamiltonian can be diagonalised exactly to
obtain the ground state energy. Again, one uses the method of `divide
and conquer'. When the six-spin sub-Hamiltonians are added together
to obtain the full Hamiltonian, the \( J_{1},J_{3},J^{\prime } \)
bonds are counted twice and the \( J_{2} \) bonds only once. One
can identify the region of parameter space in which the exact ground
state of the full ladder is of the product form. Fig. 2 shows the
phase boundaries, in the parameter space of \( \frac{J_{2}}{J_{1}} \)
and \( \frac{J^{\prime }}{J_{1}} \) for different values of \( \frac{J_{3}}{J_{1}} \)
. In the parameter regime below each phase boundary, the exact ground
state is a product over the ground states of the rungs and the tetrahedral
clusters. One finds that in certain parameter regimes \( J_{2} \)
can be larger than \( \frac{J^{\prime }}{4} \) and the exact ground
state continues to be of the product form.

We next include an external magnetic field term \( -h\sum ^{6N}_{i=1}S^{z}_{i} \)
in the Hamiltonian \( H \) (Eq. (2)), where 6N is the total number
of sites in the ladder. We first consider the case of a single tetrahedral
cluster in the presence of a magnetic field. The magnetic field couples
to the z-component of the total spin of the cluster, \( S^{z}_{tot} \),
which is a conserved quantity. The ground state energy \( E_{g}(S_{tot}^{z}) \)
at \( h=0 \) for \( S_{tot}^{z}=0,1 \) and \( 2 \) can be obtained
from Table I. When the external field \( h\neq 0 \), the ground state
in each \( S^{z}_{tot} \) subspace is \( E_{g}(S_{tot}^{z},h)=E_{g}(S_{tot}^{z},0)-hS_{tot}^{z} \).
The ground state magnetization curve can be easily obtained. Consider
the case \( J_{3}<J_{1} \). The magnetization per site \( m \) is
zero from \( h=0 \) upto a critical field \( h_{c_{1}}=J_{1} \).
For \( h_{c_{1}}<h<h_{c_{2}}=2J_{1} \), \( m=\frac{1}{4} \) and
beyond \( h=h_{c_{2}} \), the saturation magnetization, \( m=\frac{1}{2} \),
is obtained. Thus there are three magnetization plateaus at \( m=0 \),
\( \frac{1}{4} \) and \( \frac{1}{2} \). For the external field
\( h=0 \), we have already seen that there is an extended parameter
regime in which the exact ground state of the full ladder is a product
of the ground states of the rungs and the tetrahedral clusters. We
now investigate whether the same holds true for a finite magnetic
field. Again, one uses the method of `divide and conquer' and the
sub-Hamiltonian used is a six-spin cluster consisting of a tetrahedral
cluster and a rung. For the full ladder, one can identify a region
(region A) in parameter space in which for \( 0<h<h_{c_{1}} \), \( m \)
is zero. At \( h_{c_{1}} \), there is a jump in the value of \( m \)
to \( m=\frac{1}{6} \) and a plateau is obtained for \( h \) upto
\( h_{c_{2}} \) (Fig. 3). When \( h_{c_{1}}<h<h_{c_{2}} \), the
exact ground state has the tetrahedral clusters in their \( S^{z}=1 \)
ground states and the rungs in singlet spin configurations. Since,
the number of tetrahedral clusters is N and the total number of sites
is 6N, the magnetization/site \( m \) in the ground state is \( \frac{1}{6} \).
The quantization condition in Eq. (1) is obeyed as unit period of
the ground state contains six spins so that \( S_{u}=3 \) and the
magnetization \( m_{u} \) in the unit period is 1. At \( h_{c_{2}} \),
there is a second jump in \( m \) from \( \frac{1}{6} \) to \( \frac{1}{3} \).
When \( h_{c_{2}}<h<h_{c_{3}} \), the exact ground state has the
tetrahedral clusters in their \( S^{z}=2 \) ground states and the
rungs in singlet spin configurations. In this case, \( S_{u} \) and
\( m_{u} \) in Eq. (1) are 3 and 2 respectively. At \( h=h_{c_{3}} \),
there is a jump in \( m \) from \( \frac{1}{3} \) to the saturation
magnetization \( \frac{1}{2} \). For \( J_{3}<J_{1} \), \( h_{c_{1}} \),
\( h_{c_{2}} \) and \( h_{c_{3}} \) have the values \( J_{1} \),
\( 2J_{1} \) and \( J^{\prime }+J_{2} \), \( (2J_{1}<(J^{\prime }+J_{2})) \)
respectively. There are other parameter regions (regions B and C)
in the parameter space in which the full plateau structure in the
\( m \) versus \( h \) plot, as shown in Fig. 3, is not obtained.
Fig. 4 shows the phase diagram for the full ladder in a magnetic field
in the \( \frac{J^{\prime }}{J_{1}} \) vs. \( \frac{J_{3}}{J_{1}} \)
parameter space and for \( \frac{J_{2}}{J_{1}}=0.2 \). The region
A exhibits the full plateau structure in \( m \) vs. \( h \) as
shown in Fig. 3. In region B, the jump in \( m \) from 0 to \( \frac{1}{6} \)
occurs at \( h=h_{c_{1}} \) (Fig. 3) but beyond \( h_{c_{2}} \),
the ground state is no longer of the product form. In region C, the
ground state loses its simple product structure beyond \( h=h_{c_{1}} \).
Similar phase diagrams are obtained for higher values of \( \frac{J_{2}}{J_{1}} \)and
also for \( J_{3}>J_{1} \). One can generalise the ladder model shown
in Fig. 1 by assigning different coupling strengths \( J_{1} \),
\( J_{4} \) and \( J_{3} \) to the vertical, horizontal and diagonal
couplings of the tetrahedral clusters. Again, results similar to the
case \( J_{1}=J_{4} \) are obtained. With \( J_{4}=J_{3}=J_{2} \)
and \( J_{1}=J^{\prime } \), the two-chain frustrated ladder model
introduced by Bose and Gayen\( ^{18} \) is recovered. In a finite
magnetic field \( h \), the magnetization/site \( m \) vs. \( h \)
has a simple plateau structure\( ^{6} \).

Another generalisation of the ladder model shown in Fig. 1 is to replace
a tetrahedral cluster by a block of tetrahedral clusters. Fig. 5 shows
an example in which the block contains two tetrahedral clusters. The
six-spin blocks are separated by two-spin rungs. Again, one can show
that in an extended parameter regime, the ground state has the product
form in both zero and finite magnetic fields. The exact ground state
is the product of the ground states of the six-spin blocks and the
rungs. The ground state of a six-spin block is a RVB state. An extra
magnetization plateau exists for \( h_{c_{3}}<h<h_{c_{4}} \) in which
the ground state has all the six-spin blocks in their \( S^{z}=3 \)
ground state configurations and the rungs are in singlet spin configurations.
At \( h=h_{c_{4}} \), \( m \) jumps to its full saturation value.
Fig. 6 is the phase diagram similar to Fig. 4 for the full ladder
with \( \frac{J_{2}}{J_{1}}=0.2 \). In the `divide and conquer' method,
the full ladder Hamiltonian is a sum over eight-spin sub-Hamiltonians.
Each sub-Hamiltonian describes the interactions in a block of spins
consisting of two tetrahedral clusters and a single rung. In region
\( A_{1} \), the full plateau structure in \( m \) vs. \( h \)
is obtained. In regions \( B_{1} \), \( C_{1} \) and \( D_{1} \),
the ground state no longer has the product form beyond the fields
\( h_{c_{3}} \), \( h_{c_{2}} \) and \( h_{c_{1}} \) respectively.
Similar phase diagrams are obtained for higher values of \( \frac{J_{2}}{J_{1}} \).
One can generalise the ladder models shown in Figs. 1 and 6 by making
the blocks of tetrahedral clusters of bigger size (the total number
of spins in a block may be 4, 6, 8, 10..... etc.). Two-spin rungs
separate the blocks of spins. In certain parameter regimes, the exact
ground state is possibly the product of the exact ground states of
the rungs and the blocks of tetrahedral clusters. A full study of
such ladder models is yet to be undertaken.

In this paper, we have described a class of two-chain ladder models
consisting of blocks of tetrahedral clusters, containing n spins,
separated by two-spin rungs. We have specifically considered two cases:
n=4 and 6. We have shown that in an extended parameter regime, the
ground state of the ladder is a product over the ground states of
the rungs and the blocks of tetrahedral clusters. For n=4, we have
shown that the exact ground state consists of RVB spin configurations
in the tetrahedral clusters and the rungs are in singlet spin configurations.
For \( J_{3}=J_{1} \), the ground state is highly degenerate. When
\( J_{3} \) is \( <J_{1} \) (\( >J_{1} \)), the tetrahedral cluster
is in the RVB state \( \left| \psi _{RVB1}\right\rangle (\left| \psi _{RVB2}\right\rangle ) \)
and the exact ground state of the full ladder model is non-degenerate.
A notable feature of the ladder model is the presence of singlet excitations
in the triplet spin gap in certain parameter regimes. As already pointed
out in earlier references\( ^{11,12} \), the singlet energy level
\( \left| \psi _{RVB2}\right\rangle  \) of a tetrahedral cluster
crosses the triplet energy level at \( J_{3}=\frac{J_{1}}{2} \) (Table
I). Thus for \( \frac{J_{1}}{2}<J_{3}<J_{1} \), the singlet excitation
described by \( \left| \psi _{RVB2}\right\rangle  \) falls in the
triplet gap. Similarly, for \( J_{3}>J_{1} \), \( \left| \psi _{RVB2}\right\rangle  \)
is the ground state and for \( J_{1}<J_{3}<2J_{1} \), the singlet
excitation corresponding to \( \left| \psi _{RVB1}\right\rangle  \)
falls in the triplet gap. These features carry over to the case of
the full ladder model in the parameter region in which the exact ground
state can be written in a product form. The existence of singlet excitations
in the triplet spin gap is a characteristic feature of some other
AFM spin systems which include the \( S=\frac{1}{2} \) Heisenberg
antiferromagnet (HAFM) on the kagom\'{e} lattice\( ^{19} \), the
\( S=\frac{1}{2} \) HAFM on the pyrochlore lattice\( ^{20} \) and
some \( S=\frac{1}{2} \) AFM spin models on the \( \frac{1}{5} \)-depleted
square lattice\( ^{21,22} \).

The model shown in Fig. 1 can be generalised to bigger blocks of tetrahedral
clusters. (Fig. 5 shows blocks of two tetrahedral clusters). Instead
of the tetrahedral cluster shown in Fig. 1, one can also consider
a generalised tetrahedral cluster with the horizontal, vertical and
diagonal exchange interactions of different strengths. Again, in an
extended parameter regime, the exact ground state is found to be of
the product form. The ground states have the interesting structure
of islands of RVB spin configurations separated by singlet spin configurations
along the rungs. This type of exact ground state is not known for
other spin models including ladders with modulated exchange interactions.

The ladder models have also been studied in an external magnetic field
\( h \). In the parameter regime in which the ground states in the
different magnetization subspaces are of the product form, the magnetisation/site
\( m \) as a function of h exhibits plateaus (Fig. 3). The quantization
condition in Eq. (1) is obeyed at each plateau. Figs. 4 and 6 show
the phase diagrams for the ladder models of the types shown in Figs.
1 and 5. Both the phase diagrams show that there are extended regions
in parameter space in which the ground states in different magnetization
subspaces are of the product form. Kolezhuk\( ^{23} \) has studied
magnetization plateaus in a spin system consisting of strongly coupled
dimers which are again weakly coupled in a planar arrangement of zigzag
interactions. In our ladder models, we have two different type of
clusters: dimers (two-spin rungs) and tetrahedral clusters. Further
studies are needed to obtain the phase diagrams of the ladder models
in the full parameter space.

The authors thank S. Ramasesha and K. Tandon for letting them use
their Heisenberg Calculator (exact diagonalisation program) for quantum
spin systems. E. Chattopadhyay is supported by the Council of Scientific
and Industrial Research, India under sanction No. 9/15(186)/97-EMR-I.

\newpage

\subsubsection*{Figure Captions}

Fig. 1. Two-chain ladder model consisting of tetrahedral clusters
(solid lines) coupled to two-spin rungs (dashed lines). The exchange
interaction strengths are as shown in the Figure. \\
Fig. 2. Phase diagram of the ladder model (Fig.1) in the parameter
space of \( \frac{J_{2}}{J_{1}} \) and \( \frac{J^{\prime }}{J_{1}} \)
. The parameter space below a solid line corresponds to the phase
in which the exact ground state is a product over the ground states
of the rungs and the tetrahedral clusters.\\
Fig. 3. Plot of magnetization/site \( m \) versus external magnetic
field \( h \) for the two-chain ladder model shown in Fig. 1. The
plot is obtained in the parameter region in which the exact ground
states in different \( S_{tot}^{z} \) subspaces have the product
form. Two non-trivial magnetization plateaus occur at \( m=\frac{1}{6} \)
and \( m=\frac{1}{3} \) . \\
 Fig. 4. Phase diagram of the ladder model (Fig. 1) in a finite magnetic
field and in the parameter space of \( \frac{J^{\prime }}{J_{1}} \)
and \( \frac{J_{3}}{J_{1}} \) with \( \frac{J_{2}}{J_{1}}=0.2 \).
The regions A, B and C are explained in the text. \\
 Fig. 5. A two-chain spin ladder which consists of blocks of two tetrahedral
clusters coupled to two-spin rungs (dashed lines). The exchange interaction
strengths are as shown in the Figure. \\
 Fig. 6. Phase diagram of the ladder model (Fig. 5) in a finite magnetic
field and in the parameter space of \( \frac{J^{\prime }}{J_{1}} \)
and \( \frac{J_{3}}{J_{1}} \) with \( \frac{J_{2}}{J_{1}}=0.2 \).
The regions \( A_{1} \), \( B_{1} \) , \( C_{1} \) and \( D_{1} \)
are explained in the text.
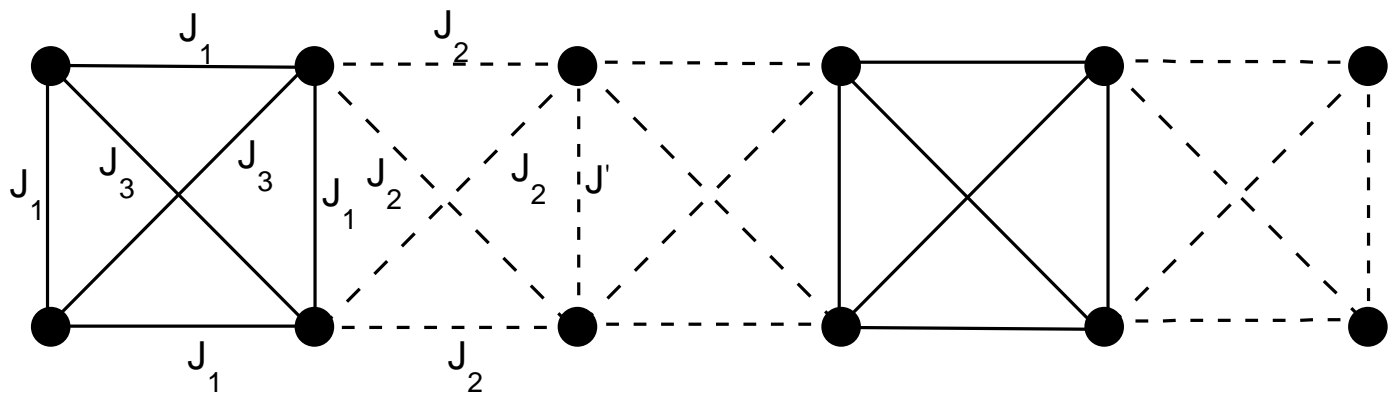
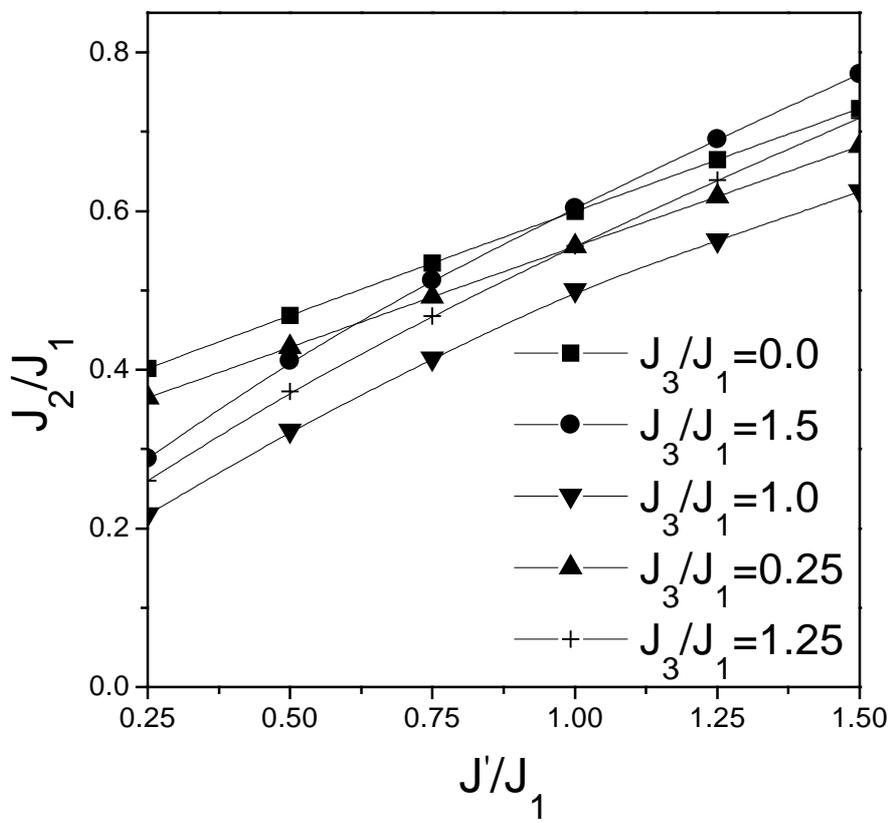
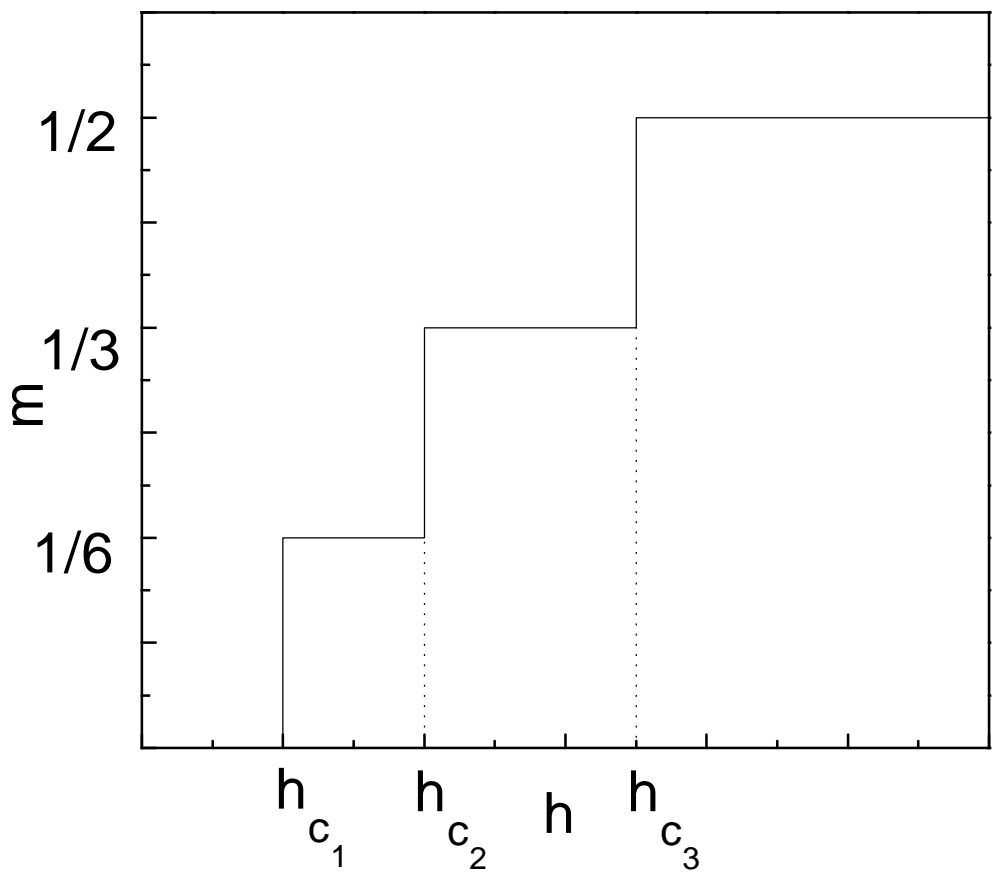
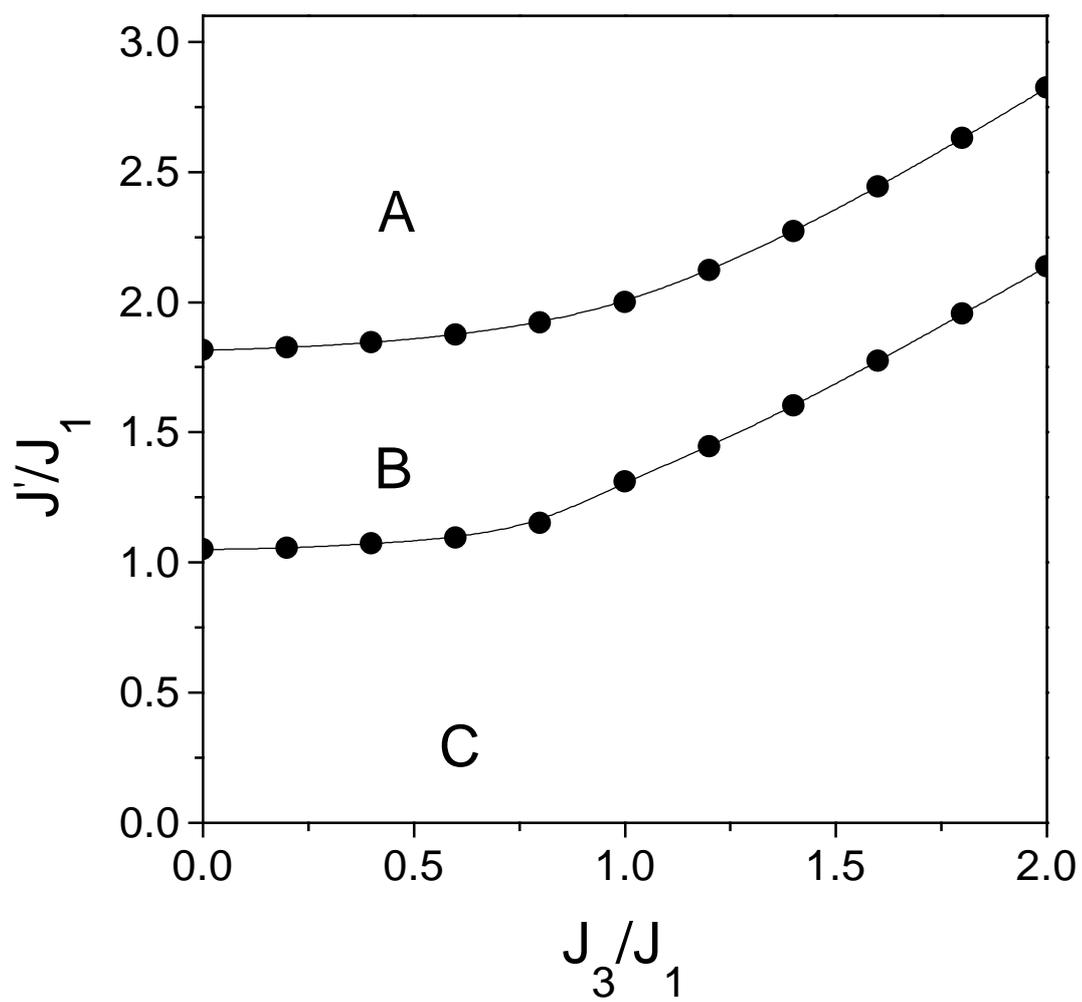
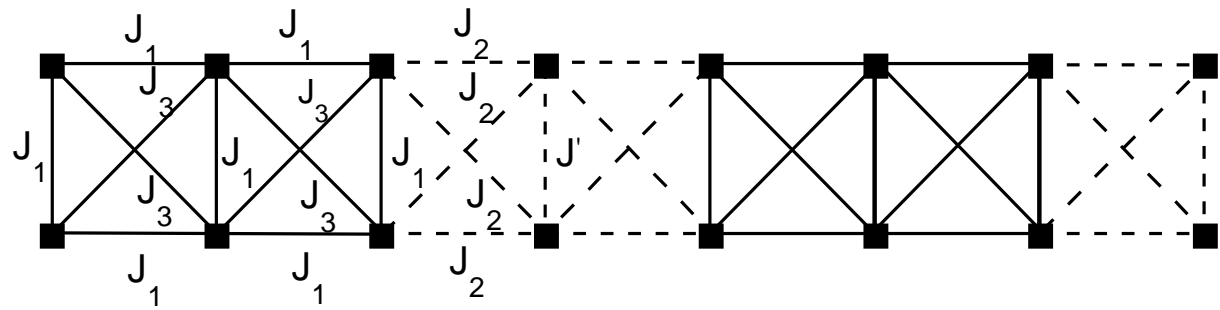
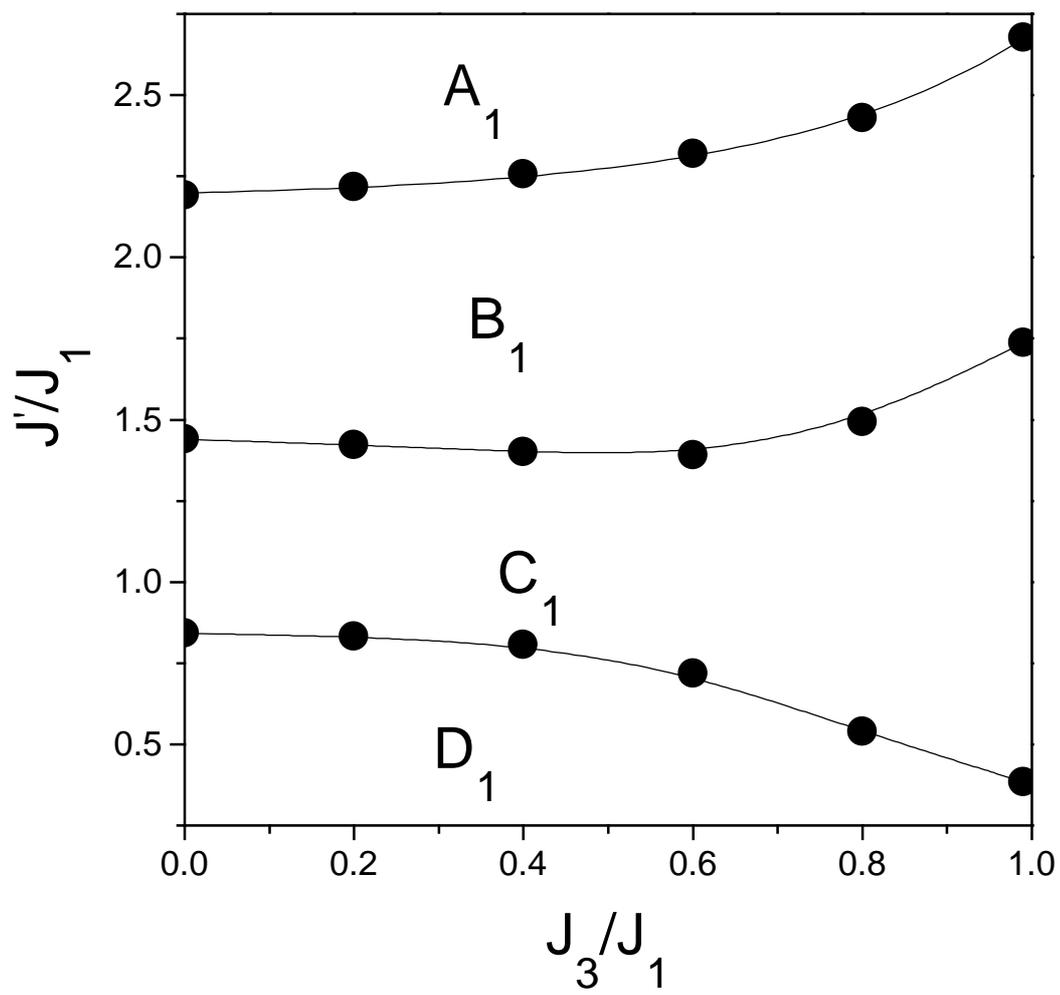
\end{document}